\documentclass[twocolumn,secnumarabic,amssymb, nobibnotes, aps, prd]{revtex4-1}

\usepackage{framed}
\usepackage{amsfonts}
\usepackage{amsmath}
\usepackage{graphicx}
\usepackage{color}

\graphicspath{{./figs/}}

\begin{document}
\title{Geometric parametric instability in  periodically modulated GRIN multimode fibers}%
\author{C. Mas Arab\'{i}, A. Kudlinski, A. Mussot, and M. Conforti }%
\affiliation{Univ. Lille, CNRS, UMR 8523-PhLAM-Physique des Lasers Atomes et Mol\'ecules, F-59000 Lille, France}
\begin{abstract}
We present a theoretical and numerical study of light propagation in  graded-index (GRIN) multimode fibers where the core diameter has been periodically modulated along the propagation direction. The additional degree of freedom represented by the modulation permits to modify the intrinsic spatiotemporal dynamics which appears in multimode fibers. More precisely, we show that modulating the core diameter at a periodicity close to the self-imaging distance allows to induce a Moir\'{e}-like pattern, which modifies the geometric parametric instability gain observed in homogeneous GRIN fibers. 
\end{abstract}
\maketitle
\section{Introduction}
Parametric resonance (PR) is a well-known instability phenomenon which occurs in systems whose parameters vary in a periodic fashion. A classical example is a pendulum whose length changes harmonically with time. Depending on the system parameters, the pendulum may be unstable and the amplitude of its oscillations may become unbounded. Another example is the formation of standing waves on the surface of a liquid enclosed by a vibrating receptacle, a process known as Faraday instability \cite{Faraday}.
Faraday instability manifests as pattern formation in several extended systems \cite{Cross93} and it has been studied in  Bose-Einstein condensates \cite{Garcia99,Staliunas2002}, granular systems \cite{Moon2001}  or chemical processes \cite{Petrov97}. In these cases, the periodic modulation of the parameters is produced by an external forcing. However, the periodic variation may be  self-induced by natural oscillations of the system as well. The emerging dynamics has been termed as geometric parametric instability (GPI) \cite{Krupa2016}. Faraday instabilities and GPI can both be observed in fiber optics, which is a particularly interesting physical system for their study due to its simplicity.
%
In nonlinear fiber optics, a continuous wave (CW) may be unstable, leading to the amplification of spectral sidebands. This effect is called modulation instability (MI) and is commonly associated to anomalous dispersion regime \cite{Agrawal}, although it is also observed in normal dispersion regime in the presence of higher order dispersion \cite{pitois2003}, fiber birefringence \cite{Wabnitz88} or multiple spatial modes \cite{Guasoni2015,Dupiol:17:2}. A CW may also be unstable whatever the dispersion regime if a parameter of the system is varied periodically (an external periodic forcing is applied), leading to the observation of PR, which can coexist with standard MI in the same system \cite{Copie2016,Conforti:16,Copie2017review,Mussot2017}. This external forcing can result from periodic amplification \cite{Matera:93}, dispersion \cite{Armaroli:12,RotaNodari2015,Finot:13,Droques2013,Droques:12} or nonlinearity \cite{Abdullaev97,Staliunas2013}. The periodic evolution of nonlinearity can be self-induced, for example, in highly multimode GRIN fibers. Indeed, such fibers exhibit a periodic self-imaging of the injected field pattern due to the interference between the different propagating modes \cite{soldano1995}. This is due to the fact that the propagation constants of the modes are equally spaced and they have almost identical group velocity \cite{Mafi2012}. This creates a periodic evolution of the spatial size of the light pattern and therefore induces a periodic evolution of the effective nonlinearity in the propagation direction  \cite{Wright2015b,Conforti2017}. 
Recent experiments showed that PRs produced by this GPI can reach detunings from the pump in the order of hundreds of THz \cite{Krupa2016}.



In the present work, we study GPI in a system having an internal and external forcings with close but not equal periodicity. We consider a multimode GRIN fiber supporting self-imaging (at the origin of the internal forcing), with an additional modulation of the core diameter (inducing the external forcing). The overall longitudinal evolution of the spatial pattern exhibits two spatial frequencies, which can induce a Moir\'{e}-like effect. This results in the generation of characteristic spectral components which differ form the usual PR frequencies.


The outline of the paper is as follows. After this Introduction, in section \ref{sec:I} we report numerical simulations  illustrating the self-imaging pattern and the parametric instability spectrum for a modulated GRIN fiber.  In section \ref{sec:II}, we calculate the CW evolution of the beam along the propagation coordinate, characterized by two spatial periods. In section \ref{sec:III}, we study the parametric instabilities generated by  the spatial pattern when group velocity dispersion is considered, and provide estimates for the frequency of unstable bands. We draw our conclusions in \ref{sec:IV}.

\section{Parametric instability in periodic GRIN fibers}\label{sec:I}

Spatiotemporal light propagation in multimode GRIN fibers can be described by the following generalized nonlinear Schr\"odinger equation (GNLSE) \cite{Krupa2016,Longhi2003}:
\begin{equation}
i\partial_zE=\frac{1}{2\beta_0}\nabla_{\bot}^2E-\frac{\beta_2}{2}\partial^2_tE-\frac{\beta_0g(z)}{2}r^2E+\chi|E|^2E,
\label{eq:GP_completa}
\end{equation}
where $E$ is the electric field envelope measured in $\sqrt{\mathrm{W}}$/m, $\nabla_{\bot}^2$ is the Laplacian over the transverse coordinates, $r^2=x^2+y^2$, $\beta_0=k_0n_0$, $n_0$ being the refractive index at the core center, $k_0$ the vacuum wavenumber at the carrier frequency $\omega_0$, $\beta_2$ is the group velocity dispersion (GVD), $g(z)=2\Delta/\rho_c^2(z)$, where $\rho_c(z)$ is the fiber core radius and $\Delta$ is the relative refractive index difference between core and cladding ($\Delta=(n_0^2-n_{clad}^2)/(2n_{0}^2)$), and $\chi=\omega_0 n_2/c$ is the nonlinear coefficient. We approximate the refractive index profile by $n^2(x,y,z)=n^2_0(1-2\Delta r^2/\rho_c(z)^2)$, so that the waveguide is modeled as pure harmonic potential with a depth varying along $z$. 
In the following, we restrict our study to light propagation over short distances (a few centimeters), so that linear coupling between modes due to fiber imperfections can be disregarded. Moreover, we assume to excite the fiber with a Gaussian beam at the center so that the field presents circular symmetry. 
 This allows to solve  Eq. (\ref{eq:GP_completa}) by means of a split-step Fourier method \cite{Agrawal} in cylindrical coordinates \cite{Guizar-Sicairos04}, significantly reducing the numerical integration time.
 
\begin{figure}[h]
\includegraphics[scale=0.9]{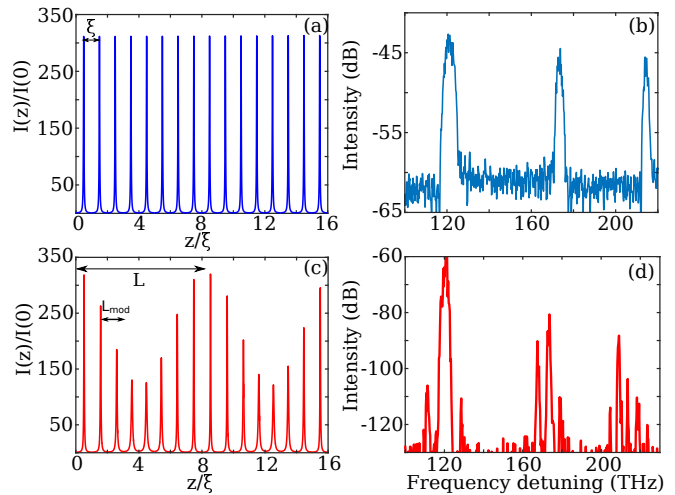}
\caption{Results obtained from direct numerical simulation of GNLSE (\ref{eq:GP_completa}). (a,c) Evolution of the intensity in the core center normalized to the input one, versus fiber length normalized to the self-imaging period $\xi$ in a fiber with (a) uniform core and (c) varying core. (b,d) Output spectrum  in a fiber with (b) uniform core and (d) varying core.  Parameters typical of commercially available GRIN fibers: $n_0=1.47$, $n_2=3.2 \times 10^{-20}$m$^2$/W, $\rho_0 = 26 ~ \mu$m, $\Delta=8.8 \times 10^{-3}$, $\lambda_0=1064$ nm, $a_0=20~\mu$m, fiber length (b) 1.4 cm (d) 3 cm, $I_0$=20 GW/cm$^2$  and $\beta_2=16.55 \cdot 10 ^{-27}$ s$^2$/m. }
\label{fig:Figura_1}
\end{figure}

We start by considering a uniform commercial GRIN fiber, whose parameters are reported in the caption of Fig. \ref{fig:Figura_1}. By injecting into this fiber a Gaussian beam of size $a_0=20~\mu$m at a wavelength  $\lambda_0=1064$ nm, a periodic light pattern is generated with a period $\xi=\pi/\sqrt{g_0}\approx 600~\mu$m ($g_0=2\Delta/\rho_0^2$). This can be observed in Fig. \ref{fig:Figura_1}(a), which shows the evolution of the intensity in the core center $I(z)=|E(z,x=0,y=0)|^2$  (normalized to the input one) as a function of fiber length (normalized to the self-imaging period $\xi$). Figure \ref{fig:Figura_1}(b) shows the output spectrum obtained at a normalized distance of 16, which exhibits three distinct GPI sidebands in the range 100 - 220 THz detuning from the pump. We then consider the propagation of the same input beam in a modulated fiber,  with a period $L_{mod}$  of 700 $\mu$m and a modulation depth of 3 $\mu m$.
Figure \ref{fig:Figura_1}(c)  shows  that the varying core diameter induces an additional modulation of the light intensity along the propagation. The self-imaging pattern is now modulated by an envelope of longer period $L$. As we will see hereafter, $L$ depends on the relation between the natural self-imaging period ($\xi$) and modulation period ($L_{mod}$). In the present example, $L_{mod}$ is chosen to be close to $\xi$ and the resulting dynamics may be understood as a kind of Moir\'e pattern. As shown in Fig. \ref{fig:Figura_1}(d), this double periodicity in the spatial behaviour produces new spectral bands around those obtained with a uniform fiber, in a similar fashion to what was observed in a single-mode dispersion-oscillating fiber with doubly periodic dispersion \cite{Copie:15}. The process leading to the generation of these new spectral components will be analyzed and explained in the following sections. 

\section{Calculation of the self-imaging pattern }\label{sec:II}

To compute the spatial evolution of the beam profile, we start from  Eq. (\ref{eq:GP_completa}) assuming CW propagation ($\partial_t=0$):
\begin{equation}
i\partial_zE=\frac{1}{2\beta_0}\nabla_{\bot}^2E-\frac{\beta_0g(z)}{2}r^2E+\chi|E|^2E.
\label{eq:GP_spacial}
\end{equation}
An approximated solution of Eq. (\ref{eq:GP_spacial})  in the weakly nonlinear regime, is a Gaussian beam with  parameters varying along the propagation coordinate, which can be calculated by exploiting the method of moments \cite{PerezGarcia2000} or variational techniques \cite{Karlsson:91}. The solution reads as:
\begin{equation}
|E_s(x,y,z)|^2=A_0^2\left(\frac{a(z)}{a_0} \right)^2\text{exp}\left(-\frac{r^2}{a(z)^2} \right), \label{eq:spacial_intensity}
\end{equation}
where $a(z)$ is the solution of the following equation (dot stands for $z$ derivative):
\begin{equation}
\ddot{a}+g(z)a+\frac{C}{a^3}=0, \quad C \equiv \left(\frac{n_2a_0^2 A_0^2}{2n_0}-\frac{1}{\beta_0^2}\right).
\label{eq:a}
\end{equation}
The whole dynamics is thus ruled by the beam radius $a(z)$. Equation (\ref{eq:a})  is a singular nonlinear Hill equation of Ermakov type whose solution can be written as \cite{Pinney1950,Carinena2008}:
\begin{equation}\label{a}
a(z)=\sqrt{u(z)^2-\frac{Cv(z)^2}{W^2}},
\end{equation}
where $u(z),v(z)$ are two linearly independent solutions of the equation:
\begin{equation}
\ddot{x}+g(z)x=0,
\label{eq:Hill_equation_x}
\end{equation}
$W=u\dot{v}-\dot{u}v=const.$ is the Wronskian, and the initial conditions are $u(0)=a_0$, $\dot{u}(0)=\dot{a}(0)$, $v(0)=0$ and $\dot{v}(0)\neq 0$. The linear Hill equation (\ref{eq:Hill_equation_x}) is solvable in closed form only for very particular forms of $g(z)$ \cite{Nayfeh1995}. 
We consider here an harmonic modulation of the fiber core $\rho(z)=\rho_0(1+\delta\cos (kz))$, where $\delta$ describes the amplitude and $k=2\pi/L_{mod}$  the  period of modulation. 

In the $(L_{mod},\delta)$ plane, there are regions (known as Arnold tongues) where solutions of Eq. (\ref{eq:Hill_equation_x}) becomes unbounded. This instability stems from the fact that the natural spatial frequency  $k_{nat}\equiv \sqrt{g_0}=\sqrt{2\Delta}/\rho_0$ of the oscillator described by Eq. (\ref{eq:Hill_equation_x}) is varied periodically with wavenumber $k$. The tips of the Arnold tongues fulfill the parametric resonance condition $k_{nat}=m\cdot k/2$ ($m$ integer), i.e. the natural spatial frequency  is multiple of half the wavenumber of the spatial forcing, which gives the following condition on the ratio between modulation period and self-imaging distance:
\begin{equation}\label{PRz}
L_{mod}=m\, \xi,\quad m=1,2,\ldots
\end{equation}
When condition (\ref{PRz}) is not satisfied, the evolution of the beam size is periodic (or quasi-periodic) \cite{Nayfeh1995}. Moreover, even if the condition (\ref{PRz}) is satisified, there is a threshold on the modulation depth $\delta$ for the emergence of parametric instability, which in general increases for higher order resonances.  For the rest of the paper we assume to be in absence of \emph{spatial} parametric resonances. 

Depending on the ratio between the modulation and self-imaging period, we can recover three different situations: $L_{mod}\ll\xi$, $L_{mod}\approx\xi$ and $L_{mod}\gg\xi$. The first case is the less interesting: in fact, the fast oscillating  terms in Eq. (\ref{eq:Hill_equation_x}) can be averaged out, and the beam evolves as in a uniform fiber. The other two cases present quite peculiar behaviors, which are analyzed in details below.

\subsection{Moir\'e pattern: $L_{mod}\approx \xi$}
\begin{figure}[t]
\includegraphics[scale=0.9]{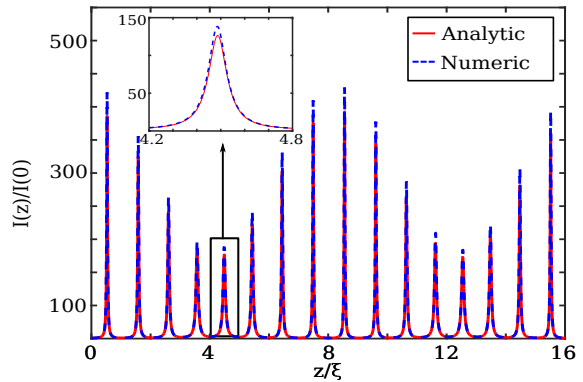}
\caption{Evolution of the intensity in the core center calculated from numerical solution of Eq. (\ref{eq:GP_spacial}) (blue curve) and exact solution (red curve) obtained from Eq. (\ref{eq:spacial_intensity},\ref{a},\ref{u}-\ref{v}). Same parameters of Fig. \ref{fig:Figura_1}(c).}
\label{comparative}
\end{figure}

The most interesting case rises when $L_{mod}\approx \xi$ and the modulation depth $\delta$ is small enough to avoid unbounded evolution.  By assuming a small modulation amplitude $\delta\ll 1$, and expanding $1/\rho(z)$ at second order in the small parameter $\delta$, Eq. (\ref{eq:Hill_equation_x}) takes the following form:
\begin{equation}
\ddot{x}+\frac{2\Delta}{\rho_0^2}(1-2\delta\cos(kz)+3\delta^2\cos^2(kz))x=0.
\label{eq:Hill_equation_x_Taylor}
\end{equation}
Two independent approximate solutions of Eq. (\ref{eq:Hill_equation_x_Taylor}) can be found by using multi-scale techniques (see appendix \ref{ap:appendix_2}):
\begin{widetext}
 \begin{align}  
\label{u} u(z)&=\frac{(4g_0-k^2)a_0}{(4+2\delta)g_0-k^2}\left(\cos(\sqrt{g_0}\sigma z)+\frac{\delta g_0}{k}\left(\frac{\cos((\sqrt{g_0}\sigma-k)z)}{2\sqrt{g_0}-k}-\frac{\cos((\sqrt{g_0}\sigma+k)z)}{2\sqrt{g_0}+k}\right)\right), \\
\label{v} v(z)&=-B\left(\sin(\sqrt{g_0}\sigma z)+\frac{\delta g_0}{k}\left(\frac{\sin((\sqrt{g_0}\sigma-k)z)}{2\sqrt{g_0}-k}-\frac{\sin((\sqrt{g_0}\sigma+k)z)}{2\sqrt{g_0}+k}\right)\right),  
 \end{align}
\end{widetext}
where
 \begin{align}
 W&=-B\frac{(4g_0-k^2)a_0}{(4+2\delta)g_0-k^2}\sqrt{g_0}\sigma, \\
 \sigma&=1+\frac{\delta^2(8g_0-3k^2)}{4(4g_0-k^2)},
 \end{align}
and $B\neq 0$ is a constant, which does not appears in the expression of $a(z)$. Equations (\ref{u}-\ref{v}) gives the expression of the beam radius through Eq. (\ref{a}). A comparison between the beam evolution obtained from Eqs. (\ref{u}-\ref{v}) and numerical solution of Eq. (\ref{eq:a}) is reported in Fig. \ref{comparative}, which shows a remarkable agreement. In general the solution is a combination of trigonometric functions of incommensurate arguments, an thus is not strictly periodic.  For small enough core radius modulation, we can safely assume $\sigma\approx 1$. In this case, the spatial behaviour can be described as a combination of functions with period $\xi$ and functions with period $L_{mod}$. If we choose $L_{mod}$  and $\xi$ to be commensurate, the evolution of the beam is periodic with a longer period which verifies $L=p L_{mod}=q\xi$ ($p,q$ integers). In this regime, the resulting spatial behavior can be understood as a Moir\'e pattern, where two patterns with close periodicity $\xi$ and $L_{mod}$ are superimposed, giving birth to a new longer periodicity  $L$. If $L_{mod}$  and $\xi$ are incommensurate, the evolution is only quasi-periodic. However, we can find $p,q$ that approximate the quasi-periodic evolution with periodic one with great accuracy. As we will illustrate below, the GPI spectrum resulting from a periodic or quasi-periodic evolution does not presents qualitative differences.

\subsection{Adiabatic modulation: $L_{mod}\gg \xi$}
\begin{figure}
\centering
\includegraphics[scale=0.9]{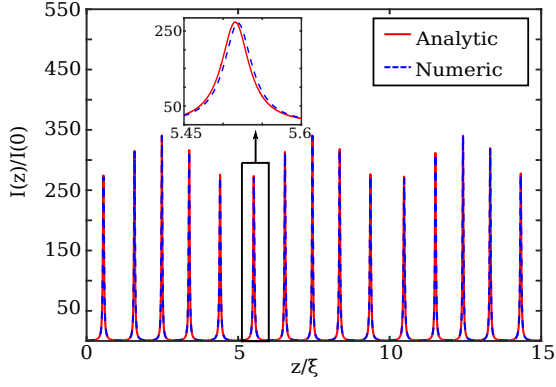}
\caption{Evolution of the intensity at core center calculated from numerical solution of Eq. (\ref{eq:a}) (red dashed line) and analytical solution Eq. (\ref{eq:adiabatic_aprox}) (blue solid line) for $L_{mod}=5\xi$.  Other parameters as in Fig. \ref{fig:Figura_1}(c). }
\label{fig:fiqure_adiabatic}
\end{figure}

 When $L_{mod}\gg \xi$, variations over the intensity are adiabatic, and we can use the Wentzel-Kramers-Brillouin (WKB) approximation \cite{Bender1999}  to find two  independent solutions $u(z),v(z)$ of Eq. (\ref{eq:Hill_equation_x}), which inserted in Eq. (\ref{a}) give the following expression for the beam size:
\begin{equation}
a(z)=\sqrt{a_0^2\sqrt{\frac{g(0)}{g(z)}}\cos^2(\phi(z))-\frac{C}{a_0^2\sqrt{g(0)g(z)}}\sin^2(\phi(z))}.
\label{eq:adiabatic_aprox}
\end{equation}
The phase $\phi(z)=\int_0^zdz'\sqrt{g(z')}$ reads as:
\begin{align}\label{phase}
\phi(z)&=\frac{2\sqrt{g_0}}{k\sqrt{1-\delta^2}}\left(\tan^{-1}\left[\sqrt{\frac{1-\delta}{1+\delta}}\tan\left(\frac{kz}{2}\right)\right]+\pi m \right),\\
\nonumber m&=\Big \lfloor \frac{z}{L_{mod}}+\frac{1}{2} \Big \rfloor,
\end{align}
where $\lfloor x \rfloor=\max \{ m \in \mathbb{Z} | m \leq x \}$.
 A comparison between the beam evolution obtained from Eqs. (\ref{eq:adiabatic_aprox}-\ref{phase}) and numerical solution of Eq. (\ref{eq:a}) is reported in Fig. \ref{fig:fiqure_adiabatic}, which shows a perfect agreement.
We can see that the fast-varying self-imaging pattern of period $\xi$ is modulated adiabatically in a sinusoidal fashion by the longer period $L_{mod}$ ($L_{mod}=5\xi$ in this example). 
  When the amplitude of modulation tends to 0 ($\delta\rightarrow 0$), the phase simplifies to $\phi(z)\approx\pi/\xi z$. In this limit, the spatial behavior can be  again described as a combination of functions with period $\xi$ and functions with period $L_{mod}$. If $L_{mod}$  and $\xi$ are chosen to be commensurate, the evolution of the beam is periodic with a longer period which verifies $L=p L_{mod}=q\xi$, like the case where $L_{mod}\approx \xi$. In the general case, where $\delta\neq 0$ we observe a self-imaging pattern modulated by an envelope of period $L=L_{mod}$, which give rise to a quasi-periodic behavior.

\section{Linear stability analysis and GPI gain}\label{sec:III}
We move now to the study of the stability of the CW spatial profile found in the previous section with respect to time periodic perturbations. We assume a spatiotemporal field of the form:
\begin{equation}\label{ansats}
E(x,y,z,t)=(1+\delta E(z,t))E_s(x,y,z),
\end{equation}
where $E_s$ is the approximated spatial electric field from Eq.  (\ref{eq:spacial_intensity}) and $\delta E(z,t)$ is a small perturbation homogeneous in the transverse plane. By substituting ansatz (\ref{ansats}) in Eq. (\ref{eq:GP_completa}), after linearization we obtain:
\begin{equation}
iE_s\partial_z\delta E=-\frac{\beta_2}{2}E_s\partial_t^2\delta E+\chi |E_s|^2E_s(\delta E+\delta E^*),
\end{equation}
which projected on $E^*_s$, gives:
\begin{equation}\label{linear}
i\partial_z\delta E=-\frac{\beta_2}{2}\partial_t^2\delta E+\frac{\chi A_0^2a_0^2}{2a(z)^2}(\delta E+\delta E^*).
\end{equation}
A similar equation can be obtained when studying MI in single mode fibers with oscillating nonlinearity \cite{Matera:93,RotaNodari2015,Abdullaev97,Armaroli:12}. By considering a time harmonic perturbation of the form $\delta E=a(z)e^{i\Omega t}+b^*(z)e^{-i\Omega t}$, we get the following system:
\begin{equation}
	\begin{bmatrix}
	\dot{a} \\
	\dot{b}
	\end{bmatrix}
	=i\begin{bmatrix}
	\frac{\beta_2\Omega^2}{2}+F(z) & F(z) \\
	-F(z) & -\left(\frac{\beta_2\Omega^2}{2}+F(z)\right) \\	
	\end{bmatrix}
	\begin{bmatrix}
	a \\
	b
	\end{bmatrix},
	\label{eq:system}
\end{equation}
where 
\begin{equation}
F(z)=\frac{\chi A_0^2a_0^2}{2a(z)^2}, 
\end{equation}
which can be reduced to a second order ordinary differential equation for $y=a+b$:
\begin{equation}
\ddot y+\frac{\beta_2\Omega^2}{2}\left(\frac{\beta_2\Omega^2}{2}+2F(z)\right)y=0.
\label{eq:Hill_equation}
\end{equation}

Since we have restricted our analysis to bounded evolutions of $a(z)$, the function $F(z)$ is (quasi) periodic as  discussed in previous section. In this case, Eq. (\ref{eq:Hill_equation}) is again a Hill equation. Note that Eq. (\ref{eq:Hill_equation}) rules the evolution of the time periodic perturbations, so that an unbounded evolution for $y$ implies the instability of the CW field. The values of the perturbation frequency $\Omega$ for which the evolution of $y$ becomes unbounded gives thus the GPI spectrum.

\subsection{Constant core diameter}

\begin{figure}
\includegraphics[scale=0.9]{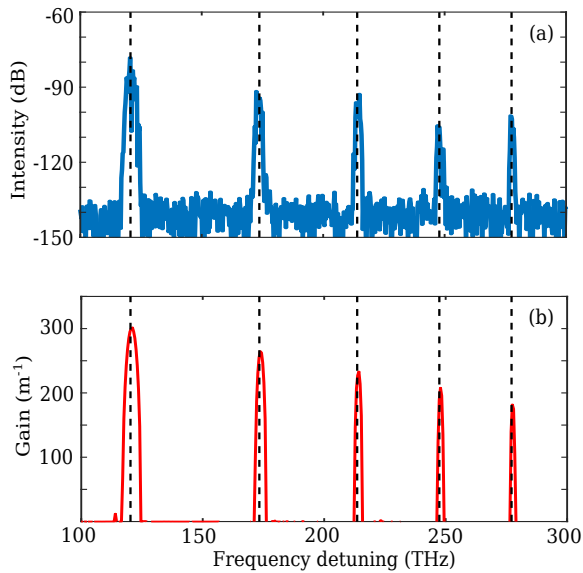}
\caption{GPI gain for the uniform fiber case. (a) Output spectrum obtained by solving numerically GNLSE (\ref{eq:GP_completa}). Black dashed lines corresponds to unstable frequencies obtained from parametric resonance condition Eq. (\ref{eq:Omega_lim}). (b) Floquet spectrum obtained from Eq. (\ref{eq:system}). Parameters: $A_0^2=$20~GW/cm$^2$, $a_0$ = 20 $\mu$m and fiber length 2.2 cm. Remaining parameters as in Fig. \ref{fig:Figura_1}(a-b).  }
\label{fig:2}
\end{figure}

We start by considering a uniform fiber, which has already been studied  in \cite{Longhi2003}. In this case the beam radius $a(z)$  assumes the following simple expression:
\begin{equation}\label{aconst}
a^2(z)=a_0^2[\cos^2(\sqrt{g_0}z)+\mathcal{C}\sin^2(\sqrt{g_0}z)],
\end{equation}
where $\mathcal{C}=(1-\mathcal{P})/(\beta_0^2a_0^4g_0)=-C/(g_0 a_0^4)$ and $\mathcal{P}=n_2\beta_0^2A_0^2a_0^2/2n_0$ is a dimensionless parameter measuring the distance from beam collapse (here we assume $\mathcal{P}\ll 1$) \cite{Karlsson:91,Longhi2003}. Expression (\ref{aconst}) is recovered by putting $\delta=0$ in Eqs. (\ref{a},\ref{u}-\ref{v}).
The average value of $F(z)$ can be calculated as:
\begin{equation}
F_{av}=\frac{1}{\xi}\int_0^\xi F(z)dz=\frac{\chi A_0^2}{2\sqrt{\mathcal{C}}}.
\end{equation}
 The parametric resonance condition for Eq. (\ref{eq:Hill_equation}), which reads:
\begin{equation}
\frac{\beta_2\Omega^2}{2}\left(\frac{\beta_2\Omega^2}{2}+2 F_{av}\right)=\left( m\,\frac{\pi}{\xi} \right)^2,
\end{equation}
permits to find the central frequencies of the GPI bands as:
\begin{equation}
\Omega_{m}^2=\frac{2}{\beta_2}\left(-F_{av}+\sqrt{F_{av}^2+g_0\,m^2} \right),\quad m=1,2,\ldots
\label{eq:Omega_lim}
\end{equation}
%
%
%
%
%
%
The full GPI spectrum can be calculated by means of Floquet theory. This consist in solving system (\ref{eq:system}) for two independent initial conditions (f.i. $[1,0]^T$ and $[0,1]^T$) over one period ($z=\xi$ in this case). This system can be solved only numerically because of the nontrivial form of $F(z)$. The two solutions calculated at $z=\xi$ compose the two columns of  the $2\times 2$ Floquet matrix, whose eigenvalues $\mu_1,\mu_2$ determine the stability of the CW solution. If one eigenvalue (say $\mu$) has modulus greater than one, the solution is unstable and the perturbations grow exponentially along $z$ with gain $G(\Omega)=\ln|\mu|/\xi$.

Figure \ref{fig:2} presents a comparison between the spectrum obtained from numerical solution of full GNLSE (\ref{eq:GP_completa}), starting from a CW perturbed by a small random noise, for 2.2 cm long uniform fiber [Fig. \ref{fig:2}(a)], and the gain spectrum obtained from Floquet theory [Fig. \ref{fig:2}(b)]. Vertical black dashed lines represent the frequencies $\Omega_{m}$ given by Eq. (\ref{eq:Omega_lim}) for $m=1,\ldots,5$, which are in excellent agreement with the locations of maximum gain. Note that in the low intensity limit ($A_0 \rightarrow 0$), they can be approximated as  $\Omega_m^2\approx 2\pi m/(\xi \beta_2)$ \cite{Krupa2016}.  The gain spectrum obtained from Floquet theory, plotted in Fig. \ref{fig:2}(b), is in excellent agreement with the spectrum obtained from direct numerical simulation, plotted in Fig. \ref{fig:2}(a),  concerning the frequency and width of the gain bands, as well as their relative intensity.

\subsection{Varying core diameter}

When the core diameter varies periodically, the spatial dynamics becomes richer and the self-imaging pattern is modulated, as described by Eqs. (\ref{a},\ref{u}-\ref{v}) for $L_{mod}\approx \xi$, or Eqs. (\ref{eq:adiabatic_aprox},\ref{phase}) for $L_{mod}\gg \xi$.  As we will show in the following, this spatial dynamics generates additional GPI bands.

\begin{figure}
\includegraphics[scale=0.9]{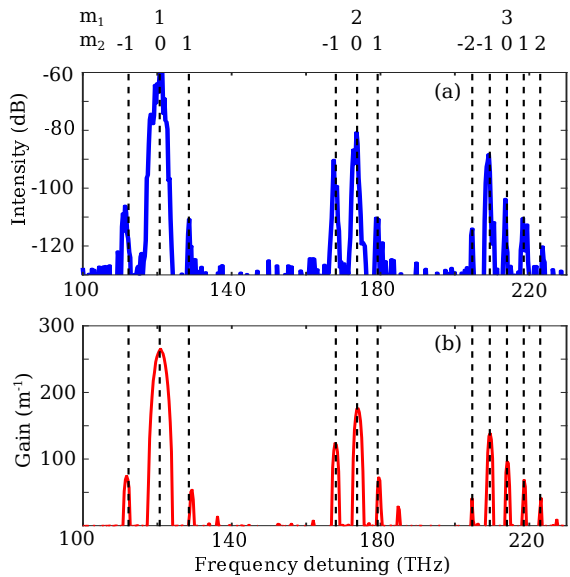}
\caption{Modulated fiber case. (a) Output spectrum obtained by solving numerically GNLSE (\ref{eq:GP_completa}). Black dashed lines corresponds to unstable frequencies obtained from Eq. (\ref{eq:Omega_lim_2}). (b) Floquet spectrum obtained from Eq. (\ref{eq:system}).  Parameters: $\delta=0.12$, $L_{mod}=(8/7)\xi$, $A_0^2=20$ GW/cm$^2$ and fiber length 3 cm. Remaining parameters as in Fig. \ref{fig:Figura_1}(a-b).  }
\label{fig:4}
\end{figure}

For the sake of simplicity, we consider a modulation period commensurate with the self imaging distance, i.e. $p\, L_{mod}=q\, \xi$ (where $p$ and $q$ are two integers). In this case the spatial evolution can be considered periodic with period $L=q\, \xi$, which allows us to obtain a simple analytic expression for the maxima of GPI gain. The parametric resonance condition for Eq. (\ref{eq:Hill_equation}) reads now:
\begin{equation}
\frac{\beta_2\Omega^2}{2}\left(\frac{\beta_2\Omega^2}{2}+2 F_{av}\right)=\left( m\,\frac{\pi}{L} \right)^2,
\end{equation}
where we have assumed that the average nonlinear forcing term  $F_{av}$ is the same as in the case of the uniform fiber. It is convenient to write the integer as $m=m_1q+m_2$, which gives the following expression for the central frequencies of the GPI bands:
\begin{equation}
\Omega_{m_1,m_2}^2=\frac{2}{\beta_2}\left(-F_{av}+\sqrt{F_{av}^2+g_0\left(m_1+\frac{m_2}{q}\right)^2} \right).
\label{eq:Omega_lim_2}
\end{equation}
The index $m_1=0,1,\ldots$ counts the main resonances, which  correspond to the ones obtained in the uniform case for $m_2=0,m_1\ne 0$ [see Eq. (\ref{eq:Omega_lim})]. The most important information which can be extracted from Eq. (\ref{eq:Omega_lim_2}) is that we have the generation of new sidebands around the principal ones.
The number $m_2$ describes these additional sub-harmonic  resonances deriving from the longer modulation  period $L>\xi$. The interval of values for $m_2$ depends on the parity of $q$, and is defined formally for $q$ even as:
\begin{equation}
q=2n \Rightarrow \left\{\begin{array}{rcc}
-n+1 & \le m_2 \le n,&  m_1\ne 0,\\
   1  &\le m_2 \le n,&  m_1= 0,
\end{array}\right. 
\end{equation}
or odd as:
\begin{equation}
q=2n+1 \Rightarrow \left\{\begin{array}{rcc}
-n & \le m_2 \le n,&  m_1\ne 0,\\
   1  &\le m_2 \le n,&  m_1= 0.
\end{array}\right. 
\end{equation}
For example we consider now a shallow modulation ($\delta=0.12$) of the fiber, with a period $L_{mod}=8/7\xi$. Figure \ref{fig:4}(a) reports the spectrum after a propagation of 3 cm from numerical simulations of GNLSE (\ref{eq:GP_completa}). We can observe the characteristic splitting of bands due to the double periodicity. Black dashed lines show the frequency of unstable sidebands given by Eq. (\ref{eq:Omega_lim}),  which are in excellent agreement with the numerical simulation result.  We can calculate the width and position of bands by performing a numerical Floquet analysis as described in the previous section. As in the constant core case, the frequency, width and relative intensity of the sidebands is in excellent agreement with numerical simulations (Fig. \ref{fig:4}(a)).

\begin{figure*}
\includegraphics[scale=0.9]{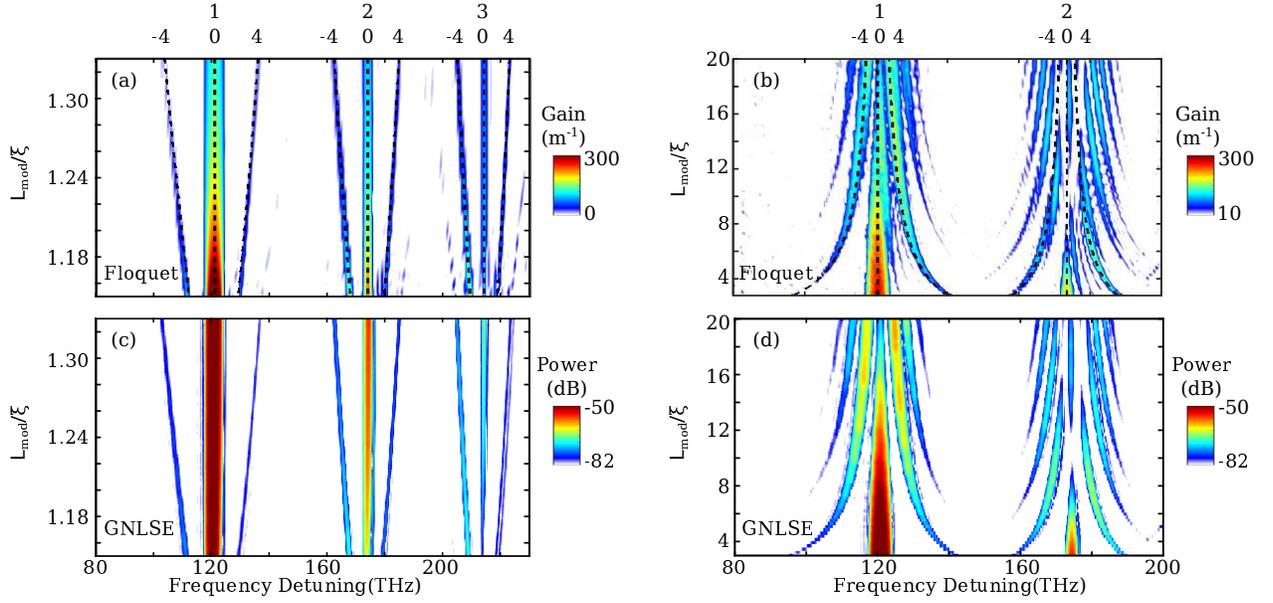}
\caption{(a,b) Gain map obtained with numerical Floquet analysis as a function of the ratio $L_{mod}/\xi$, for (a) $L_{mod}\approx \xi$ and (b) $L_{mod} \gg \xi$. (c,d) Direct numerical integration of equation \ref{eq:GP_spacial}, for (c) $L_{mod}\approx \xi$ and (d) $L_{mod} \gg \xi$. Parameters: $\beta_2=16.55 \times 10^{-27}$ s$^2$/m, $a_0=20~\mu$m , $\rho_0=26~\mu$m and $A_0^2=20~GW/cm^2$ }
\label{fig:figure3}
\end{figure*}

It is interesting to investigate the the gain spectrum as a function of the ratio $L_{mod}/\xi$ between the modulation period and the self-imaging distance. We start by considering the range $L_{mod}\approx \xi$, where the spatial Moir\'e pattern is generated. Figure \ref{fig:figure3}(a) shows the GPI gain map as a function of frequency and modulation period, calculated by means of the Floquet theory described before, using expressions Eqs. (\ref{a},\ref{u}-\ref{v}) for the beam size. The corresponding ratio between periods has been chosen to be commensurate and verifies $L_{mod}=(n+16)/(n+12) \xi$, where $n$ is an integer in the interval $[0,20]$. The overall period of the spatial pattern is thus given by $L=(n+16)\xi$.
A remarkable agreement is found with the unstable frequencies predicted by Eq. (\ref{eq:Omega_lim}), which are reported in Fig. \ref{fig:figure3}(a) as black dashed curves. For each principal resonance $m_1=1,2,3$, one can notice the generation of additional sideband pairs $m_2=\pm 4$, whose frequency separation increases with the modulation period. These additional sidebands pairs appear at index $m_2=\pm 4$ because, for the definition of the modulation period used in this example, each value of $n$ generates an overall period $L$ composed four slow oscillations (the pattern is quasi-periodic of period $\approx L/4$).

The other interesting case occurs when the modulation period of the core diameter is chosen to be several times greater than self-imaging distance, i.e. $L_{mod}\gg \xi$, where the spatial pattern is described by  Eqs. (\ref{eq:adiabatic_aprox},\ref{phase}). In this case, the relation between $L_{mod}$ and $\xi$ has been chosen to verify $L_{mod}=(12+n)/4\xi$, where now $n$ is taken in the interval $[0,68]$, giving an overall period $L=(n+12)\xi$. The corresponding Floquet gain map is reported in \ref{fig:figure3}(b), where one can notice some qualitative differences. For the smallest modulation period, only a pair of sidebands appears, well detached from the fundamentals ones, corresponding to $m_1=1,2$. By increasing the modulation period, additional sidebands pairs stem out, and they all tends to cluster around the principal ones. For the longer modulation period considered here, at least three couples of sidebands merge with the fundamental ones, giving rise to a single structured band.
Also in this case, a remarkable agreement is found with the unstable frequencies predicted by Eq. (\ref{eq:Omega_lim}) (black dashed curves). 
  
The results of Floquet analysis are essentially supported by direct numerical integration of Eq. (\ref{eq:GP_completa}), as evidenced by   Figs. \ref{fig:figure3}(c) and (d), which reports the output spectrum after a propagation distance of 3 cm as function of $L_{mod}$. In order to get rid of any randomness in the initial condition, we add a coherent seed to the CW (a short hyperbolic secant of duration 1 fs and intensity 10$^{-5}$ smaller than the cw pump).  Moreover, $L_{mod}$ has been chosen to be equally spaced between the upper and lower limits considered, so the ratio $L_{mod}$ and $\xi$ are not necessary commensurate. This fact confirms that considering the spatial behavior as strictly periodic does not produce any qualitative difference.

\section{Conclusions}\label{sec:IV}
We have theoretically studied GPI in a graded-index multimode fiber with an axially-modulated core diameter. We have shown that a periodic modulation of the core diameter 
allows to generate new GPI sidebands. 
We have developed a theory to predict the frequency of these additional sidebands, which is in excellent agreement with direct numerical simulations of the GNLSE and Floquet stability analysis. In our simulations we have used realistic parameters, this fact gives a solid evidence the described effects can be experimentally observed. This study contributes to a further  understanding  of  the  rich  dynamics related to nonlinear waves propagating in  multimode fibers.

\section*{Acknowledgements}
The authors acknowledge discussions with G. Martinelli and O. Vanvincq. This work has been partially supported by IRCICA, by the Agence Nationale de la Recherche through the NOAWE (ANR-14-ACHN-0014) project, the LABEX CEMPI (ANR-11-LABX-0007) and the Equipex Flux (ANR-11-EQPX-0017 projects), by the Ministry of Higher Education and Research, Hauts-de-France Regional Council and European Regional Development Fund (ERDF) through the CPER Photonics for Society P4S.

\appendix

%

\section{Multiscale analysis}
\label{ap:appendix_2}
In this appendix, we give some details on the method used to approximate the beam evolution in the limit $L_{mod}\approx \xi $. We start from Eq. (\ref{eq:Hill_equation_x_Taylor}), where we define the dimensionless coefficients $w=\sqrt{g}/k$, $\epsilon=w^2\delta$ and $\bar{z}=kz$, to get:
\begin{equation}
\ddot{x}+\left(w^2-2\epsilon \cos(\bar{z})+\frac{3\epsilon^2}{w^2}\cos^2(\bar{z})\right)x=0.
\label{eq:apendix_2_2}
\end{equation}
The initial assumption $L_{mod}\approx \xi$  implies $w\approx 1$, thus we can safely perform a multiscale development in  $\epsilon\ll 1$ \cite{Nayfeh1995} up to first order. By defining $\bar{z}_n=\epsilon^n\bar{z}$ and $x=\sum_{n=0}\epsilon^nx_n$, and equating equal powers of $\epsilon$, an infinite heriarchy of equations is obtained. From this infinite set equations, we retain only up to order $\epsilon^2$: 

\begin{align}
\epsilon^0: \quad (D^2_0+w^2)x_0&=0, \label{eq:B2}\\
\epsilon^1: \quad (D^2_0+w^2)x_1&=-2(D_1D_0-\cos(\bar{z}_0))x_0,  \label{eq:B3}\\
\epsilon^2: \quad (D^2_0+w^2)x_2&=-2(D_1D_0-\cos(\bar{z}_0))x_1- \label{eq:B4}\\
&-(2D_2D_0+D_1^2+\frac{3}{2w^3}\cos(\bar{z}_0))x_0, \nonumber
\end{align}

where $D_n=\partial_{\bar{z_n}}$. At order $\epsilon^0$, we find:
\begin{equation}
x_0=A(\bar{z}_1,\bar{z}_2)e^{iwz}+A^*(\bar{z}_1,\bar{z}_2)e^{-iwz} \label{eq:B5}
\end{equation}

where $A(\bar{z}_1,\bar{z}_2)$ is a complex function depending on the slower variables $\bar{z}_1,\bar{z}_2$. To know  the dependence of this fuction on  $z_1$ and $z_2$, Eq. (\ref{eq:B5}) is substituted in Eq. (\ref{eq:B3}) and secular terms are imposed to vanish. We obtain the following solution for $x_1$:
\begin{equation}
x_1=A(\bar{z}_2)\left(\frac{-e^{i(w+1)\bar{z}_0}}{2w+1}+\frac{e^{i(w-1)\bar{z}_0}}{2w-1}\right)+c.c
\end{equation}
where c.c denotes complex conjugate. To find $A(\bar{z}_2)$, we substitute $x_1$ in Eq. (\ref{eq:B4}) and impose again secular terms to vanish, this bring us to:
\begin{equation}
A=\frac{C}{2}e^{i(\phi \bar{z}_2+\beta)} \quad , \quad \phi=\frac{8w^2-3}{4w^3(4w^2-1)},
\end{equation}
where $C$ and $\beta$ are two real constants fixed by the boundary conditions. By writing the complete function $x=x_0+\epsilon x_1+O(\epsilon^2)$ we obtain the general solution:
\begin{widetext}
\begin{equation}
x=C\left[\cos(\tilde{w}z+\beta)+\epsilon\left(\frac{\cos((\tilde{w}-1)z+\beta)}{2w-1}-\frac{\cos((\tilde{w}+1)z+\beta)}{2w+1}\right)\right]+O(\epsilon^2)~,~\tilde{w}=w\left(1+\frac{\delta^2(8\omega^2-3)}{4(4\omega^2-1)}\right)
\end{equation}
\end{widetext}

From this equation, we can readily obtain $u(z)$ an $v(z)$ by imposing the appropriate boundary conditions.

\bibliographystyle{ieeetr}
\bibliography{multimode}

\end{document}